%% file: final.tex
\documentclass[amsmath,  floatfix, nofootinbib, prx, aps, usenames, dvipsnames,10pt, allowfontchageintitle, a4paper,twocolumn]{quantumarticle}
\usepackage{algorithm}
\usepackage{algpseudocode}
\usepackage{float}
\usepackage[skip=2pt]{caption}
\pdfoutput=1
\usepackage[utf8]{inputenc}
\usepackage[english]{babel}
\usepackage[T1]{fontenc}
\usepackage{amsmath}
\usepackage{hyperref}

\usepackage{tikz}
\usepackage{lipsum}
\input{definitions.tex} 
\usepackage{soul}

\begin{document}

\title{Variational measurement-based quantum computation for generative modeling}

\author{Arunava Majumder}
\email{Arunava.Majumder@uibk.ac.at}
\affiliation{Institute for Theoretical Physics, University of Innsbruck, Technikerstr. 21a, A-6020 Innsbruck, Austria} 
\author{Marius Krumm}
\affiliation{Institute for Theoretical Physics, University of Innsbruck, Technikerstr. 21a, A-6020 Innsbruck, Austria} 
\author{Tina Radkohl}
\affiliation{Institute for Theoretical Physics, University of Innsbruck, Technikerstr. 21a, A-6020 Innsbruck, Austria} 
\author{Lukas J. Fiderer}
\affiliation{Institute for Theoretical Physics, University of Innsbruck, Technikerstr. 21a, A-6020 Innsbruck, Austria} 
\author{Hendrik Poulsen Nautrup}
\affiliation{Institute for Theoretical Physics, University of Innsbruck, Technikerstr. 21a, A-6020 Innsbruck, Austria} 
\author{Sofiene Jerbi}
\affiliation{Dahlem Center for Complex Quantum Systems, Freie Universität Berlin, Berlin, Germany} 
\author{Hans J.~Briegel}
\affiliation{Institute for Theoretical Physics, University of Innsbruck, Technikerstr. 21a, A-6020 Innsbruck, Austria}

\date{}
\maketitle

%
%
\onecolumn

\begin{abstract}\vspace{-0.75cm}
Measurement-based quantum computation (MBQC) offers a fundamentally unique paradigm to design quantum algorithms. Indeed, due to the inherent randomness of quantum measurements, the natural operations in MBQC are not deterministic and unitary, but are rather augmented with probabilistic byproducts. Yet, the main algorithmic use of MBQC so far has been to completely counteract this probabilistic nature in order to simulate unitary computations expressed in the circuit model. In this work, we propose designing MBQC algorithms that embrace this inherent randomness and treat the random byproducts in MBQC as a resource for computation. As a natural application where randomness can be beneficial, we consider generative modeling, a task in machine learning centered around generating complex probability distributions. To address this task, we propose a variational MBQC algorithm equipped with control parameters that allow one to directly adjust the degree of randomness to be admitted in the computation. Our algebraic and numerical findings indicate that this additional randomness can lead to significant gains in expressivity and learning performance for certain generative modeling tasks, respectively. These results highlight the potential advantages in exploiting the inherent randomness of MBQC and motivate further research into MBQC-based algorithms.
\end{abstract}\vspace{-0.6cm}

\section{Introduction}

Among the different models of quantum computation, the circuit model is arguably the most popular model for designing quantum algorithms \cite{nielsen2010quantum}. In this model, quantum computation is performed mainly in a deterministic fashion by following a sequence of unitary gates, and the only randomness in the computation arises from the measurements performed as a final step. In that regard, measurement-based (or one-way) quantum computation (MBQC) \cite{briegel2009m,raussendorf2001a,Raussendorf2003cluster} stands out as a radically different paradigm for quantum computation. In this model, one instead starts by a preparing large, entangled state (commonly a cluster state \cite{briegel2001persistent}) which is universal for quantum computing, and the actual computation is then entirely performed by a sequence of single-qubit measurements acting on this state. Since measurements have inherently random outcomes, this model naturally gives rise to a probabilistic (non-unitary) quantum computation. Notoriously however, it is possible to simulate the standard circuit model using MBQC \cite{Raussendorf2003cluster}. This requires imposing an adaptive measurement scheme that corrects for random measurement outcomes as to simulate a deterministic unitary computation. In practice, this is also how MBQC is most commonly used for quantum computation: algorithms are natively designed in the circuit model and the MBQC implementation simply simulates these quantum circuits via adaptive measurements.

In this work, we go beyond the standard use of MBQC and propose seeing its inherent randomness not as a hurdle to overcome, but as a feature to be exploited. We suggest only partially correcting random measurements, in a way that is also specified by the MBQC algorithm. This results in a controllable type of randomness to be left in the quantum computation that is natural to implement in the MBQC model but artificial to simulate in the circuit model.

It is not \textit{a priori} clear which computational tasks could truly benefit from this inherent randomness. A good candidate however can be found in the field of machine learning. In generative modeling \cite{Tomczak2022}, one is indeed given the task of generating new samples from a certain target probability distribution, given only access to a limited number of training samples from this distribution. Naturally, this computational task requires some degree of randomness or ``tailored noise'' to generate good candidate distributions. A quantum algorithmic approach for generative modeling that has been gaining traction in recent years is based on so-called quantum circuit born machines (QCBM) \cite{liu2018d, benedetti2019generative}. In this type of variational quantum algorithm \cite{cerezo2021v}, distributions generated by variational quantum circuits are taken as candidate distributions for a generative modeling task and are trained via classical optimization algorithms to approximate the target distribution.

In this paper, we explore the idea of designing QCBMs in the MBQC model. Conceptually, our proposed model allows one to adjust the degree of adaptiveness in an MBQC computation, and therefore tailor the randomness of the computation to best fit potential target distributions. We develop the framework for designing MBQC computations with controllable adaptiveness and showcase both numerically and algebraically its advantage over fully adaptive MBQC in generative learning tasks. Specifically, we consider a variational MBQC model with $XY$-plane measurements on regular lattices which corresponds to the circuit model computation with probabilistic Pauli operators as in Fig.~\ref{fig:architecture}~\cite{nautrup2023measurement}.

This manuscript is structured as follows. In Sec.~\ref{rel_work}, we relate our approach to previous works in the literature. In Sec.~\ref{methods}, we present the MBQC framework and our newly introduced variational MBQC model. Sec.~\ref{res} is dedicated to the numerical and algebraic investigation of different design choices for variational MBQC and exploring a potential learning advantage in generative modeling. In Sec.~\ref{conclusion}, we present our conclusions and outlook.

\section{Related works}\label{rel_work}

Quantum computers hold the promise to solve some classes of problems more efficiently than their classical counterparts. In their seminal work \cite{bravyi2018q}, Bravyi \emph{et al.} established an unconditional quantum advantage for constant-depth quantum circuits over constant-depth classical circuits in a relational task they coined the $2D$ hidden linear function (HLF) problem. Interestingly, the quantum circuits that solve 2D-HLF are naturally expressed as a \emph{non-adaptive} MBQC computation on a cluster state \cite{briegel2001persistent}, where the correlations between the random outcomes of MBQC are key to finding correct solutions. This result therefore showcases an interesting use of non-adaptive MBQC. A related result that also establishes a quantum advantage using non-adaptive MBQC is that of Miller \emph{et al.} \cite{miller2017q}. In this work, the authors showed that constant-depth MBQC can generate distributions that are inaccessible to polynomial-depth classical circuits, under a widely-believed conjecture in complexity theory (the non-collapse of the polynomial hierarchy). Primordially, the random byproducts resulting from the non-adaptive MBQC computation contribute to the randomness of the probability distribution generated, in a way that allows it to be generated in constant depth, while preserving its quantum advantage.

The main application for our variational MBQC model that we explore in this work is that of generative modeling. In the classical literature, a wide array of generative models have been proposed, each characterized by their unique architecture, strengths, and limitations. Among these models,  large language models \cite{radford2019language}, latent diffusion models \cite{rombach2022high}, and variational auto-encoders \cite{kingma2013auto} are some of the most notable ones. These models are used for various tasks, including text and data generation, data augmentation, and anomaly detection \cite{schlegl2017unsupervised}. The models specialize in unraveling the intricate connections within high-dimensional target distributions, a particularly demanding task because generative modeling often entails working with limited information access.  Quantum generative learning models are a class of generative models that leverage quantum mechanics to enhance the learning process. Some of the popular examples are generative adversarial networks \cite{goodfellow2014generative} and quantum circuit born machines \cite{liu2018d}.  Similar to classical generative models, their primary goal is to approximate an unknown target state or distribution.

In a recent work \cite{ferguson2021measurement}, Ferguson \emph{et al.}~already introduced a measurement-based model for variational quantum computation, which they called the measurement-based variational quantum eigensolver. This model comes in two variants. The first approach consists of simulating circuit-based \textit{Ansätze} with MBQC and aims for more efficient resource utilization. This aspect was further investigated in a follow-up work \cite{qin2023a}. The second approach proposes augmentations to MBQC computations that have no direct analog in the circuit model and generate a broader range of variational state families. These findings, while highlighting the potential of MBQC, differ from our approach as they do not exploit the \textit{inherent randomness of MBQC} to go beyond the circuit model. In another recent study \cite{wu2023r}, Wu \emph{et al.}~demonstrated that injecting random unitaries with trainable probabilities into variational quantum circuits could significantly enhance their expressivity and had the potential to greatly boost their training performance. These findings, while obtained in supervised learning tasks, are supportive of an advantage in designing probabilistic variational quantum computations. Nonetheless, the approach of Wu \emph{et al.}~does not exploit the inherent randomness of MBQC but rather injects externally generated random unitaries in the circuit model. 

%
%

\section{Methods}\label{methods}

%
%
\begin{figure*}
    \centering
    \includegraphics[width=\linewidth]{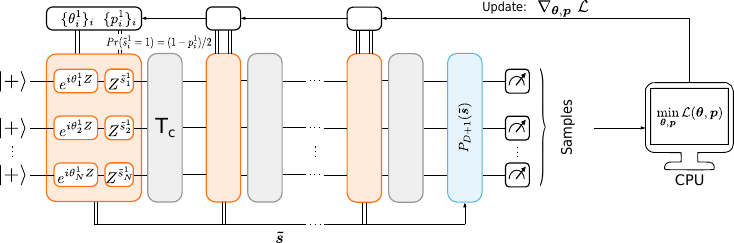}
    \caption{\small{\textbf{Variational measurement-based quantum computation architecture.} Depicted is the $N$-qubit circuit model picture for the variational MBQC proposed in this paper. Here, MBQC can be understood as $D$ alternating layers of local $R_Z$-rotations (parametrized by $\{\theta^j_i\}_{i,j}$), byproduct operators $Z^{\tilde{s}^j_i}$ and Clifford Quantum Cellular Automata $T_c$ (C-QCA) as exemplified in Fig.~\ref{fig:glider}. In this approach to generative modeling, we learn the variational angles $\{\theta^j_i\}_{i,j}$ and probabilities $\{p^j_i\}_{i,j}$ for correcting byproducts. To better control the impact of the randomness on the computation, we also maintain conditional control of Pauli operators at the end of the computation on the appearance of byproducts by $P_{D+1}(\bm{\tilde{s}})$ [see Eq.~\eqref{eq:mbqc_byproducts}] as in standard adaptive MBQC. As shown in Sec.~\ref{sec:two_models_compare}, this significantly improves the expressivity and versatility of the learning model. The resulting mixed unitary channel is given by Eq.~\eqref{eq:mbqc_channel_model}. An implicit loss is calculated classically by comparing measurement samples from the MBQC with samples from the real distribution.}}
    \label{fig:architecture}
\end{figure*}

\subsection{PQC, generative models, and QCBM}\label{pqc,gm}

A parametrized quantum circuit (PQC) is a specialized quantum circuit composed of single and two-qubit gates equipped with adjustable parameters. Over time, researchers have put forth various strategies employing PQCs to tackle generative learning tasks \cite{benedetti2019generative, liu2018d}. The efficient sampling capabilities inherent in quantum circuits serve as a compelling motivation for us to delve deeper into discerning the properties of PQCs that can boost their effectiveness in generative learning tasks. In this article, the PQC that we will be using is inspired and prescribed by MBQC and hence referred to as variational MBQC. A schematic diagram of this variational MBQC can be found in Fig.~\ref{fig:mbqc-circuit-equi} where the quantum circuit is referred to as a PQC.
\vspace{8pt}

Generative learning refers to a category of unsupervised machine learning methods that aim to model the underlying probability distribution of a data set. In other words, generative models learn how data is generated and can subsequently generate new data samples that resemble the original data set. Generative models can be of two types. \textit{Explicit} generative models provide a tractable expression for the probability distribution $q_{\bm{\theta}}(x)$ they generate ($\bm{\theta}$ being the variational parameters and $x$ being a sample), while \textit{implicit} models define a stochastic process that generates samples from this distribution \cite{mohamed2016learning}. We will mainly talk about \textit{implicit} generative models in this paper. The training data set of a generative model is a collection of $M$ independent and identically distributed ($i.i.d.$) samples $D=\{ x_{1}, x_{2}, .., x_{M} \}$, where the samples $x_{k}\in \{ 0,1 \}^{N}$ are $N$-bit strings. The corresponding distribution $\pi(x)$ is unknown and the goal is to design and train a model that can approximate $\pi(x)$ by generating its own samples from its empirical model distribution $q_{\bm{\theta}}(x)$. 
\vspace{8pt}

The key ingredient of classical generative models is a neural network (NN), with trainable weights and biases, that can learn an underlying probability distribution. In contrast, quantum generative modeling replaces the NN with a quantum circuit, and the overall model here is referred to as QCBM. The foundational structure of a QCBM closely resembles that of the VQE algorithm \cite{peruzzo2014variational}, which comprises several essential components: a sufficiently expressive PQC incorporating both single-qubit parametrized gates and two-qubit gates, a classical post-processing stage where we compute the cost function and its corresponding gradient using samples generated by the PQC, and a feedback loop that iteratively updates the PQC parameters based on the cost function until convergence is achieved. This cyclic process continues until the cost function reaches a stable state. In a standard QCBM, a PQC prepares a parametrized state $|\psi(\bm{\theta})\rangle$, and a quantum measurement following the Born rule $q_{\bm{\theta}}(x)=| \langle x|\psi(\bm{\theta})\rangle |^{2}$ is used to generate the model distribution. Each run of the PQC produces a single sample from $q_{\bm{\theta}}(x)$. The fundamental advantage of a QCBM is that it allows efficient unbiased sampling from the encoded distribution which is a desirable property in generative models.

\subsection{Measurement-based quantum computation}\label{mbqc}
\begin{figure}[t]
    \centering
    \includegraphics[width=\linewidth]{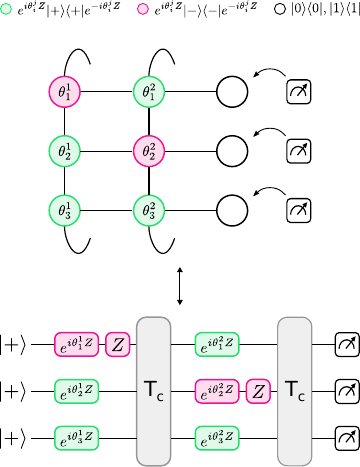}
    \caption{\small{\textbf{Circuit picture of an MBQC based on C-QCA.} An MBQC on the cluster state (top) which only utilizes measurements in the $XY$-plane can be understood in the circuit picture (bottom) where local $Z$-rotations are interleaved with C-QCAs $T_c$. The output qubits in the third column (white) are measured in the $Z$-basis. Measurements on a cluster state are inherently random. In this example the bottom left qubit of the cluster state is measured in the $e^{i\theta_3^1 Z}\ket{\pm}\bra{\pm}e^{-i\theta_3^1 Z}$-basis, yielding the outcome $+1$ and thus showing no $Z$-byproduct in the circuit. We also depict two qubits (pink, top left and center) for which the measurements yielded an outcome $-1$, resulting in $Z$-byproducts in the equivalent circuit. These byproducts can be corrected to make the computation unitary, but we will implement such corrections only with a certain probability.
    }}
    \label{fig:mbqc-circuit-equi}
\end{figure}
%

MBQC is a model for universal quantum computation that is driven by onsite measurements on an entangled \emph{resource state}~\cite{raussendorf2001a,briegel2001persistent}. In this scheme computation proceeds typically as follows: We start by initializing a so-called \emph{cluster state}~\cite{briegel2001persistent} by placing $\ket{+}$-states on nodes of a rectangular lattice and applying controlled-$Z$ ($CZ$-) gates between neighboring nodes, i.e., those connected by an edge [see Fig.~\ref{fig:mbqc-circuit-equi} (top)]. Information processing on a cluster state can be done by measuring each qubit of the cluster (from left to right) in the $XY$-plane of the Bloch sphere~\cite{Mantri2017universality}. That is, depending on the quantum gate that we wish to implement, we pick a measurement basis $\ket{\pm_{\theta^{j}_i}} := \ket{0} \pm e^{i\theta^{j}_i} \ket{1}$ for each qubit at position $(i,j)\in [N]\times [D]$ on the rectangular grid, where $[N]=\{1,2,...,N\}$ and $N\times (D+1)$ is the size of the grid, where we added one additional layer as output qubits [see Fig.~\ref{fig:mbqc-circuit-equi} (top)]. Typically, this rightmost layer is then measured in the $Z$-basis. To simplify our calculations below, we assume periodic boundary conditions such that there is an edge between $(1,j)$ and $(N,j)$ for all $j\in[D]$.

Given a cluster state and measurement angles, the deterministic unitary computation assumes that one measures $\ket{+_{\theta^{j}_i}}$ everywhere. However, it is known that for each individual qubit in the cluster state, the probability of the outcome $+1$ is exactly $\frac{1}{2}$, i.e., equally likely as measuring $\ket{-_{\theta^{j}_i}}$. Fortunately, on the cluster state, one can correct unintended outcomes $-1$ and therefore obtain a unitary computation, despite this inherent randomness: an outcome $-1$ corresponds to applying a projector $\ket{-_{\theta^{j}_i}}\bra{-_{\theta^{j}_i}}$ instead of $\ket{+_{\theta^{j}_i}}\bra{+_{\theta^{j}_i}}$. Using $\ket{-_{\theta}}\bra{-_\theta} = Z \ket{+_\theta}\bra{+_\theta} Z$, this can be understood as if one had projected onto $\ket{+}\bra{+}$ and then applied an additional $Z$-gate. Such accidental gates are referred to as \emph{byproducts}. To understand how to correct for byproducts, consider the so-called \emph{stabilizer} symmetries of a cluster state: cluster states are invariant under application of gates of the form $X_{(i,j)} \bigotimes_{k \in \mathcal{N}(i,j)} Z_k$ where $\mathcal{N}(i,j)$ are the neighbors of the qubit at $(i,j)$. Whenever a byproduct operator appears, we can compensate for it by completing a stabilizer. For example, measuring $-1$ at $(i,j)$ (corresponding to a $Z_{(i,j)}$ byproduct), we apply $X_{(i,j+1)}Z_{(i-1,j+1)}Z_{(i+1,j+1)}Z_{(i,j+2)}$, thereby making it a symmetry of the cluster state.

As it was shown in Refs.~\cite{nautrup2023measurement,stephen2019subsystem,raussendorf2019computationally,raussendorf2005quantum}, an MBQC can be understood in the circuit model where a $Z$-rotated measurement $e^{i\theta^j_i Z}\ket{\pm}_{(i,j)}\bra{\pm}e^{-i\theta^j_i Z}$ at position $(i,j)$ in the grid, becomes a local $Z$-rotation $e^{i\theta^j_i Z_i}$ (parameterized by an angle $\theta^j_i$) on qubit $i$ at depth $j$ interleaved with a certain Clifford quantum cellular automaton (C-QCA) [see Fig.~\ref{fig:mbqc-circuit-equi} (bottom)]. 
In general, C-QCAs are locality-preserving, translationally invariant Clifford operations, typically acting on a ring of qubits (i.e., assuming periodic boundary conditions). 
In the case of standard MBQC on a cluster state, the corresponding C-QCA, $T_c$, is a product of $CZ$-gates and Hadamard gates (see Fig.~\ref{fig:glider}). As C-QCAs map Clifford gates onto Clifford gates, they are fully determined by their commutation relation with Pauli-operators. That is, we write,
\begin{equation}
\label{eq:T-func_3}
    \begin{split}
        T_c(X_i) &:= T_cX_iT^\dagger_c = X_{i-1} Z_i X_{i+1} \\
        T_c(Z_i) &:= T_cZ_iT^\dagger_c = X_i 
    \end{split}
\end{equation}
for all $i\in [N]$ (which holds because we assumed periodic boundary conditions). The corresponding locality-preserving, translational invariant Clifford circuit is depicted in Fig.~\ref{fig:glider}.  As detailed in Ref.~\cite{nautrup2023measurement} the (universal) family of gates, parameterized by $\bm{\theta}=\{\theta_i^j\}_{i,j}$, that is implementable by this circuit {Fig.~\ref{fig:architecture} (assuming no byproducts), is given by
\begin{align}\label{eq:mbqc_unitary}
    U_c(\bm{\theta})&=\prod_{j=D,...,1}\left( T_c\prod_{i=N,...,1}\exp\left(i\theta^{j}_iZ_i\right)\right)\nonumber\\
    &=\left(\prod_{\substack{j=D,...,1\\i=N,...,1}}\exp\left({i\theta^{j}_iT_c^{D-j+1}(Z_i)}\right)\right)\cdot T^D_c
\end{align}
where $T_c^k(Z_i)$ is applying $T_c$ $k$ times to $Z_i$. The first line of Eq.~\eqref{eq:mbqc_unitary} can be directly put into correspondence with the circuit in Fig.~ \ref{fig:architecture} (fixing $\boldsymbol{\tilde{s}}=\boldsymbol{0})$. The second line can be derived simply from propagating all $Z$-rotations through the circuit in Fig.~\ref{fig:mbqc-circuit-equi} to the end, in accordance with Eq.~\eqref{eq:T-func_3}. An important property of $T_c$ is that it is periodic, i.e, $T_c^L=\mathbb{I}$ for $L=N$ if $N$ is even and $L=2N$ if $N$ is odd. That is, we can simplify $T_c^a=T_c^{[a]_L}$ where $[a]_L:=a\mod L$. 

In this way, we can see how MBQC on a cluster state gives rise to an \textit{\textit{Ansatz}} for PQCs according to Eq.~\eqref{eq:mbqc_unitary} where $T_c$ is the fixed part of the \textit{\textit{Ansatz}} and measurement angles are variational parameters~\cite{nautrup2023measurement}.

Byproducts appear in this picture as additional $Z$-operators at specific positions within the circuit [see Fig.~\ref{fig:mbqc-circuit-equi} (pink)].

\begin{figure}[t]
    \centering
    \includegraphics[width=.8\linewidth]{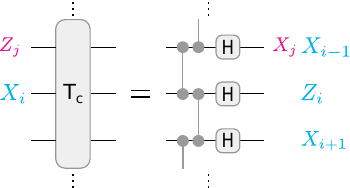}
    \centering
    \caption{ \small{\textbf{C-QCA and byproduct propagation.} The C-QCA consists of a layer of controlled-phase gates between neighboring qubits and a layer of Hadamard gates on each qubit (assuming periodic boundary conditions).  Byproduct propagation is exemplified for Pauli-$X$ and -$Z$ operators, and can be straightforwardly extended to any product of Pauli operators (see Ref.~\cite{nautrup2023measurement}). The rule for byproduct propagation can be understood by the transition function in Eq.~\eqref{eq:T-func_3}. 
    }} 
    \label{fig:glider}
\end{figure}

To formulate an expression as in Eq.~\eqref{eq:mbqc_unitary}, but including byproducts, consider a set of numbers $\bm{s}:=\{s_i^j\}_{i,j}\in\{0,1\}^{N\times D}$ that indicates the presence of a byproduct operator on qubit $i$ at layer $j$ as $Z_i^{s_i^j}$. Then, the full unitary family implemented by an MBQC given byproducts becomes,
\begin{align}\label{eq:mbqc_unitary_byprod}
    &U_c(\bm{\theta}, \bm{s})=\\&\left(\prod_{\substack{j=D,...,1\\i=N,...,1}}
    T_c^{D-j+1}(Z^{s_i^j}_i)\cdot
    \exp\left({i\theta^{j}_iT_c^{D-j+1}(Z_i)}\right)\right)\cdot T_c^D.\nonumber
\end{align}
Here, we used the same rules for propagating byproducts as with the rotations before. Interestingly, as the byproducts are simply products of Pauli-operators, we can further propagate all byproducts to the end of the computation:
\begin{align}\label{eq:mbqc_unitary_prop}
    U_c&(\bm{\theta}, \bm{s})=
    \prod_{\substack{j=D,...,1\\i=N,...,1}}T_c^{D-j+1}(Z^{s_i^j}_i)\ \cdot\\
    &\left(\prod_{\substack{j=D,...,1\\i=N,...,1}}
    \exp\left(i\theta^{j}_i\left[P_j(\bm{s}) T_c^{D-j+1}(Z_i) P_j^\dagger(\bm{s})\right]\right)\right)\cdot T_c^D\nonumber
\end{align}
where
\begin{align}\label{eq:mbqc_byproducts}
    P_j(\bm{s}):=\prod_{\substack{j'=j-1,...,1\\i'=N,...,1}}T_c^{D-j'+1}(Z_{i'}^{s^{j'}_{i'}}).
\end{align}
Since Eq.~\eqref{eq:mbqc_byproducts} are simply Pauli operators, the terms $P_j(\bm{s}) T_c^{D-j+1}(Z_i) P_j^\dagger (\bm{s})$ effectively only introduce angle flips due to the commutation relations $XZ=-ZX$. 
That is, we can write
\begin{align}\label{eq:mbqc_angle_flips}
    P_j(\bm{s})T_c^{D-j+1}(Z_i)=(-1)^{\kappa_i^j(\bm{s})}T_c^{D-j+1}(Z_i)P_j(\bm{s}),
\end{align}
for some $\kappa_i^j(\bm{s})\in\{0,1\}$ that depends on the byproducts $\bm{s}$ and position $(i,j)$ in the circuit.
%

%
%

\subsection{Variational MBQC with probabilistic byproduct correction}\label{partially corrected}

In this section, we are going to describe an MBQC-based framework for PQCs (see Fig.~\ref{fig:architecture}) corresponding to a mixed unitary channel.
For easier visualization, we will use the equivalent circuit picture of an MBQC based on C-QCA (see Sec.~\ref{mbqc}). We note, however, that this PQC is particularly well suited for an MBQC-based architecture because it does not assume any additional control beyond what is already required by an MBQC. 

Besides using a C-QCA as a circuit \textit{Ansatz} for PQC, the central element of our architecture is a proper control and utilization of byproducts in $U_c$ of Eq.~\eqref{eq:mbqc_unitary_prop}. 
Typically, random byproduct operators of an MBQC would be fully corrected to ensure that the MBQC implements a deterministic (i.e., unitary) quantum computation (see Eq.~\eqref{eq:mbqc_unitary}). In generative modeling however, it may be advantageous to allow for some (controlled) randomness. This is where our \textit{Ansatz} for variational MBQC comes in: Instead of correcting every byproduct, we introduce a \emph{correction probability} $p^j_i$ that specifies the probability with which a potential byproduct on qubit $i$ at layer $j$ is corrected. This correction probability is connected to a learnable parameter $\zeta_i^j\in\mathbbm{R}$ as 
\begin{align}\label{eq:prob_variation}
    p^j_i=\sigma(\zeta_i^j)
\end{align}
where $\sigma(\cdot)$ is the sigmoid function.
As part of the architecture in Fig.~\ref{fig:architecture}, we sample a value $c^{i}_j\in\{0,1\}$ from these probabilities as follows,
\begin{equation}
    c^j_i =
\left\{
	\begin{array}{ll}
		0  & \mbox{with  probability } 1- p^{j}_{i},\\
		1 & \mbox{with probability } p^{j}_{i}
	\end{array}
\right.
\end{equation}
Now consider the random measurement outcome $s^{j}_i\in\{0,1\}$ for a qubit $(i,j)$ in an MBQC [corresponding to a $Z^{s^{j}_i}$ byproduct in Fig. \ref{fig:mbqc-circuit-equi} (bottom)]. If $c^{j}_i=1$, we correct the byproduct, while we do not correct it for $c^{j}_i=0$. Effectively, this introduces a byproduct $Z^{\tilde{s}^{j}_i}$ in the circuit (see Fig.~\ref{fig:architecture}) where $\tilde{s}^{j}_i=(1-c^{j}_i)s^{j}_i$ $\forall i,j$. Equivalently, as $p({s^j_i=1})=\frac{1}{2}$ and by the law of total probability, we find,
\begin{align}\label{eq:sample_prob}
    p(\tilde{s}^j_i = 1) = \frac{1}{2} (1-p^j_i).
\end{align}

Naively, we could stop here and use the byproducts alone as source of randomness and rely on the family of unitaries implemented by $U_c(\bm{\theta}, \bm{\tilde{s}})$ in Eq.~\eqref{eq:mbqc_unitary_prop}. Given the learned probabilities $\bm{p}=\{p^j_i\}_{i,j}$, this would give rise to a \emph{variational MBQC as a quantum channel} as follows,
\begin{align}\label{eq:mbqc_channel_not_model}
    \mathcal{E}_c(\bm{\theta}, \bm{p})[\rho]=\sum_{\tilde{s}\in\{0,1\}^{N\times D}}
    p(\bm{\tilde{s}}) U_c(\bm{\theta}, \bm{\tilde{s}})\rho U_c^\dagger(\bm{\theta}, \bm{\tilde{s}}).
\end{align}

However, as we will see in Sec.~\ref{sec:two_models_compare}, byproducts introduce so much noise into the system that the output quickly becomes a uniform distribution with decreasing $p^j_i$. In the extreme case of discarding the measurement outcome altogether (i.e., $p^j_i=0$), this has the same effect as tracing the corresponding qubit from the cluster state by virtue of the no-signaling principle. To mitigate this detrimental impact of byproducts, we introduce Pauli-operators at the end of the computation which are conditioned on the appearance of byproducts (see Fig.~\ref{fig:architecture}). These Pauli operators are $P_{D+1}(\bm{\tilde{s}})$ (see Eq.~\eqref{eq:mbqc_byproducts}) and are designed to correct for the first term in Eq.~\eqref{eq:mbqc_unitary_prop}.
Then, the corresponding family of unitaries that is implementable by this circuit is,
\begin{align}\label{eq:mbqc_unitary_model}
    \tilde{U}_c(\bm{\theta}&, \bm{\tilde{s}})=\\
    &\left(\prod_{\substack{j=D,...,1\\i=N,...,1}}
    \exp\left((-1)^{\kappa^j_i(\bm{\tilde{s}})}i\theta^{j}_i T_c^{D-j+1}(Z_i)\right)\right)\cdot T_c^D,\nonumber
\end{align}
where we made use of Eq.~\eqref{eq:mbqc_angle_flips} to express the exponent in Eq.~\eqref{eq:mbqc_unitary_prop} through angle flips.
Given the learned probabilities $\bm{p}=\{p^j_i\}_{i,j}$, we can then define another \emph{variational MBQC as a quantum channel} as follows,
\begin{align}\label{eq:mbqc_channel_model}
    \tilde{\mathcal{E}}_c(\bm{\theta}, \bm{p})[\rho]=\sum_{\tilde{s}\in\{0,1\}^{N\times D}}
    p(\bm{\tilde{s}}) \tilde{U}_c(\bm{\theta}, \bm{\tilde{s}})\rho \tilde{U}_c^\dagger(\bm{\theta}, \bm{\tilde{s}})
\end{align}
where $p(\bm{\tilde{s}})$ is calculated as in Eq.~\eqref{eq:sample_prob}. Compared to Eq.~\ref{eq:mbqc_channel_not_model}, the model in Eq.~\ref{eq:mbqc_channel_model}, removes the classical, i.e., Clifford part of the byproducts, while retaining the non-Clifford part in form of angle flips.

In this way, our model for variational MBQC corresponds to a mixed unitary \textit{Ansatz} based on MBQC where rotation angles $\bm{\theta}$ are variational parameters that may be flipped probabilistically depending on other variational parameters $\bm{p}$ (or equivalently $\bm{z}$, see Eq.~\eqref{eq:prob_variation}). As we will see in Sec.~\ref{sec:two_models_compare}, $\tilde{\mathcal{E}}_c(\bm{\theta}, \bm{p})$ of Eq.~\eqref{eq:mbqc_channel_model} has a practical advantage over $\mathcal{E}_c(\bm{\theta}, \bm{p})$ of Eq.~\eqref{eq:mbqc_channel_not_model}.



\subsection{Cost and gradient}\label{sec:grad}


For our training purposes, we employ the Maximum Mean Discrepancy (MMD) \cite{gretton2012k} as an implicit loss function, which is defined as follows:

\begin{equation}\label{mmd loss}
\begin{aligned}
    \mathcal{L}(\bm{\theta}, \bm{p})= {} & \mathop{\mathbb{E}}_{\substack{x \sim q_{(\bm{\theta}, \bm{p})}\\ y \sim q_{(\bm{\theta}, \bm{p})}}} [K(x,y)]-2\mathop{\mathbb{E}}_{\substack{x \sim q_{(\bm{\theta}, \bm{p})}\\ y \sim \pi}} [K(x,y)] \\
    & +\mathop{\mathbb{E}}_{\substack{x \sim \pi\\ y \sim \pi}} [K(x,y)]
\end{aligned}
\end{equation}
Here, the $q_{(\bm{\theta}, \bm{p})}$ represents the distribution from our variational MBQC model with two different types of parameters $(\bm{\theta}, \bm{p})$ (see Fig.~\ref{fig:architecture}) and $\pi$ is the target distribution. The $K(x,y)$ in Eq.~\eqref{mmd loss} is referred to as the kernel function between two datapoints $x$ and $y$. For more details about the loss function, we refer to the Appendix \ref{cost_detail}

The resulting gradient for the variational  measurement angles (see Ref.~\cite{liu2018d}) of the PQC is
\begin{equation}\label{grad theta}
\begin{aligned}
    \frac{\partial \mathcal{L}}{\partial \theta^{j}_{i}} = {} & \mathop{\mathbb{E}}_{\substack{x \sim q_{(\bm{\theta^{+}},\bm{p})}\\ y \sim q_{\phi}}} [K(x,y)]-\mathop{\mathbb{E}}_{\substack{x \sim q_{(\bm{\theta^{-}},\bm{p})}\\ y \sim q_{\phi}}} [K(x,y)] \\
      & -\mathop{\mathbb{E}}_{\substack{x \sim q_{(\bm{\theta^{+}},\bm{p})}\\ y \sim \pi}} [K(x,y)]+\mathop{\mathbb{E}}_{\substack{x \sim q_{(\bm{\theta^{-}},\bm{p})}\\ y \sim \pi}} [K(x,y)]
\end{aligned}
\end{equation}
Here, $q_{(\bm{\theta^{+}}, \bm{p})}$ and $q_{(\bm{\theta^{-}}, \bm{p})}$ are the output distributions of the QCBM with parameters $\bm{\theta^{\pm}} = \bm{\theta} \pm \frac{\pi}{2} \bm{e} ^{j}_{i}$ and $\bm{e}^{j}_{i}$ is a vector with a $1$ at position $(i,j)$ and $0$s elsewhere. We do this parameter shifting for each angle individually with other angles unchanged. The parameters $\bm{\phi}$ is the set of unchanged parameters, i.e., $\bm{\phi}=(\bm{\theta},\bm{p})$.
\vspace{8pt}

The gradient for the correction probabilities takes the following form
\begin{equation}\label{gradp}
\begin{aligned}
    \frac{\partial \mathcal{L}}{\partial p^{j}_{i}} = {} & 2(\mathop{\mathbb{E}}_{\substack{x \sim q_{\bm{\phi}^{1}}\\ y \sim q_{\bm{\phi}}}} [K(x,y)]-\mathop{\mathbb{E}}_{\substack{x \sim q_{\bm{\phi}^{0}}\\ y \sim q_{\bm{\phi}}}} [K(x,y)]) \\
  & -2(\mathop{\mathbb{E}}_{\substack{x \sim q_{\bm{\phi}^{1}}\\ y \sim \pi}} [K(x,y)]-\mathop{\mathbb{E}}_{\substack{x \sim q_{\bm{\phi}^{0}}\\ y \sim \pi}} [K(x,y)])
\end{aligned}
\end{equation}
Here, $\bm{\phi}^{1}=(\bm{\theta},\bm{p}^{1})$ and $\bm{\phi}^{0}=(\bm{\theta},\bm{p}^{0})$, where 
\begin{equation}
    (p^{b})^{j'}_{i'} :=
    \begin{cases}
    p^{j'}_{i'} & \text{ for } i',j' \neq i,j\\
    b & \text{ for } i',j' = i,j
    \end{cases}
\end{equation}
for the $i'$-th qubit and $j'$-th layer, and $b\in\{0,1\}$, meaning that the original correction probabilities $p^{j'}_{i'}$ are only modified at the $(i,j)$-th location, where the correction probability is set to $0$ for $b=0$ and $1$ for $b=1$. The derivation of this gradient is deferred to Appendix \ref{gradient correction}.

\section{Results}\label{res}

%
%

In this section, we investigate the use of the variational MBQC as defined in the previous sections as a quantum generative model. 
In Secs.~\ref{sec:learn_model} and~\ref{sec:gauss}, we first demonstrate numerically the potential advantage of using a mixed unitary channel model $\tilde{\mathcal{E}}_c(\bm{\theta},\bm{p})$ [see Eq.~\eqref{eq:mbqc_channel_model}] over a deterministic (unitary) model $\tilde{\mathcal{E}}_c(\bm{\theta},\bm{p}=\bm{1})[\rho]\equiv U_c(\bm{\theta})\rho U^\dagger_c(\bm{\theta})$ [comparing Eq.~\eqref{eq:mbqc_unitary} with Eq.~\eqref{eq:mbqc_unitary_prop} for $\bm{s}=0$]. This is done for two generative learning tasks.
In Sec.~\ref{sec:two_models_compare_alg}, we provide algebraic evidence in terms of Theorem~\ref{theorem:expressivity} that probability distributions generated from the mixed unitary channel $\mathcal{E}_c(\bm{\theta},\bm{p})$ can be more expressive than those generated from $U_c(\bm{\theta})$.
In Sec.~\ref{sec:two_models_compare}, we demonstrate numerically that the model $\tilde{\mathcal{E}}_c$ in Eq.~\eqref{eq:mbqc_channel_model} offers an advantage over the model $\mathcal{E}_c$ in Eq.~\eqref{eq:mbqc_channel_not_model}. The code that we used to perform these numerical
experiments can be found on GitHub \cite{Majumder_VMBQC}.

\subsection{Learning mixed unitary channel distributions}\label{sec:learn_model}

In this section, we numerically demonstrate the advantage in expressiveness of a generative model based on mixed unitary channels $\tilde{\mathcal{E}}_c(\bm{\theta},\bm{p})$ over a unitary model $U_c(\bm{\theta})$. To this end, we randomly pick a $\tilde{\mathcal{E}}_c(\bm{\theta}_t,\bm{p}_t)$ as a target, and then numerically compare the learning performance of two models $U_c(\bm{\theta})$ and $\tilde{\mathcal{E}}_c(\bm{\theta},\bm{p})$ at the same depth and width. The target model is initialized with angles $\bm{\theta}_t$ and probabilities $\bm{p}_t$ drawn uniformly at random from $[0.8,1]$. The generated outcome distribution of the target model $\tilde{\mathcal{E}}_c(\bm{\theta}_t,\bm{p}_t)$ serves as the target distribution that the two models $U_c(\bm{\theta})$ and $\tilde{\mathcal{E}}_c(\bm{\theta},\bm{p})$ are supposed to learn. 
For updating the model parameters, we opt for the Adagrad optimizer \cite{duchi2011adaptive} and compute gradients as outlined in Eqs.~\ref{grad theta} and~\ref{gradp}. The $\bm{\theta}$ parameters of the trained models are initialized at random from a uniform distribution within the interval $[0, 1)$. Similarly, the correction probabilities $\bm{p}$ are randomly sampled from a uniform distribution within the range $[r, 1]$, where $r = 1 - 3 / (N \times D)$. As before, $N$ is the number of qubits and $D$ is the circuit depth. 
It is generally favorable to initialize the correction probabilities close to $1$ as the distribution becomes otherwise too uniform. To update both the variational angles $\bm{\theta}$ and the correction probabilities $\bm{p}$, we tried out different learning rates ($LR$) for both of them. We did a hyperparameter optimization over various learning rates within the range of $[0.01, 1]$ for both $\bm{\theta}$ and $\bm{p}$.

In our training of two different models, we always optimize the loss function corresponding to the unitary model first, with a suitable learning rate for the $\theta$ parameters ($LR_{\theta}$) only. Secondly, we use the same $LR_{\theta}$ along with the learning rate for the correction probabilities ($LR_{p}$) to optimize the loss function corresponding to the channel model. Here we tried a few $LR_{p}$s, along with fixed $LR_{\theta}$ from the unitary model. In that sense, we always try to maximize the performance of the unitary model first and later by appropriately choosing $LR_{p}$ we optimize our channel model.

We generally found that larger system sizes benefit from a larger learning rate. As an example, in Fig. \ref{fig:gauss}, for the 8 qubit simulations we choose $LR_p=0.4$ for the probabilities $\bm{p}$ and $LR_{\theta}=0.1$ for the $\bm{\theta}$ angles whereas for the 5 qubit cases, $LR_{\theta, p}=0.1$ for both $\bm{p}$ and $\bm{\theta}$. These learning rates are not optimal as our hyperparameter optimization was very simple. Different hyperparameter optimization techniques, such as a grid search, can be used to optimize the loss function of these models further. However, we do not expect that this will significantly change the conclusion from this section and therefore, consider hyperparameter optimization beyond the scope of this manuscript. 

\begin{figure}[t]
    \centering
    \includegraphics[width=8cm, height=6.5cm]{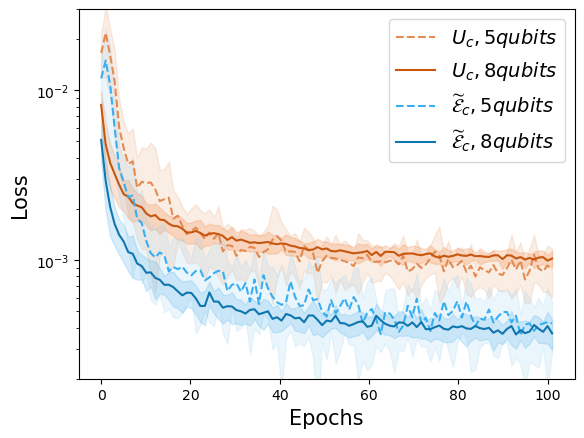}
    \caption{\small{\textbf{Variational MBQC learning performance across various system sizes.} Two learning models, based on $\tilde{\mathcal{E}}_c(\bm{\theta},\bm{p})$ (blue) in Eq.~\eqref{eq:mbqc_channel_model} and $U_c(\bm{\theta})$ (orange) in Eq.~\eqref{eq:mbqc_unitary}, respectively, are trained to approximate a given target distribution, generated from a mixed unitary channel based on $\tilde{\mathcal{E}}_c$ in Eq.~\eqref{eq:mbqc_channel_model}. All models use the same number of qubits and layers. The plot shows the mean loss across $100$ epochs averaged over 12 randomly initialized iterations. The shaded areas correspond to the respective standard deviations. \textit{Solid lines:} Learning performance of $U_c$ (orange) and $\tilde{\mathcal{E}}_c$ (blue), respectively, for $N=8$ qubits and depth $D=7$. \textit{Dashed lines:} Learning performance of $U_c$ (orange) and $\tilde{\mathcal{E}}_c$ (blue), respectively, for $N=5$ qubits and depth $D=4$. 
    This plot shows the improved performance of the mixed unitary models over the unitary models in this task.}}%
    \label{fig:main one}
\end{figure}

In Fig.~\ref{fig:main one} we show the results of the generative learning process. We use model $\tilde{\mathcal{E}}_c$ to generate the target distribution for two system sizes: $N=5$ and $N=8$ qubits, with respective circuit depths of $D=4$ and $D=7$. The training dataset comprises $5000$ samples for the $N=5$ qubits example and $10,000$ samples for the $N=8$ qubits case. Both learning models, $U_c$ and $\tilde{\mathcal{E}}_c$, undergo training with identical hyperparameters, including number of qubits, circuit depth, and dataset size. Our training utilizes the implicit loss function in Eq.~\eqref{mmd loss}, where each model learns for 100 epochs. To ensure robustness and consistency, we repeat the training process with random initialization, calculating an average loss across 12 iterations.

In Fig.~\ref{fig:main one}, we see in the averaged learning curves that there are significant gaps between the final losses achieved by the unitary models $U_c$ on the one hand and the mixed unitary models $\tilde{\mathcal{E}}_c$ on the other hand. These gaps demonstrate that mixed unitary models can be better at learning to represent mixed unitary targets than unitary models. 

\subsection{Learning double Gaussian distributions}\label{sec:gauss}

\begin{figure}[t]
    \centering
    \includegraphics[width=8cm, height=6.5cm]{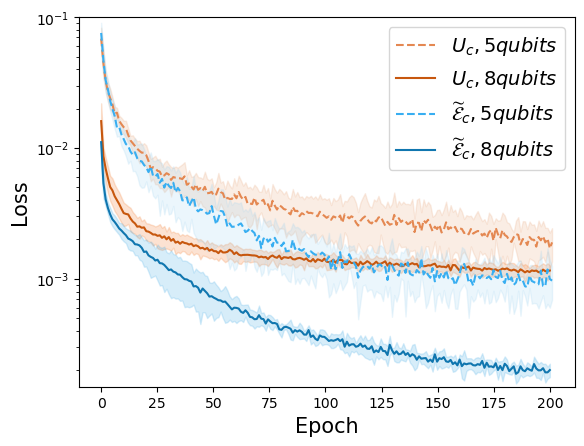}
    \caption{\small{\textbf{ Variational MBQC learning performance on a double Gaussian across various system sizes.} 
    Two learning models, based on $\tilde{\mathcal{E}}_c(\bm{\theta},\bm{p})$ (blue) in Eq.~\eqref{eq:mbqc_channel_model} and $U_c(\bm{\theta})$ (orange) in Eq.~\eqref{eq:mbqc_unitary}, respectively, are trained to approximate a given target distribution, generated from a double Gaussian distribution in Eq.~\eqref{eq:double_gauss}.
    The plot shows the mean loss over 200 episodes averaged over eight randomly initialized iterations. The shaded areas correspond to the respective standard deviations.
    \textit{Solid lines:} Learning performance of $U_c$ (orange) and $\tilde{\mathcal{E}}_c$ (blue), respectively, for $N=8$ qubits and depth $D=7$. \textit{Dashed lines:} Learning performance of $U_c$ (orange) and $\tilde{\mathcal{E}}_c$ (blue), respectively, for $N=5$ qubits and depth $D=4$. 
    This plot shows the improved performance of the mixed unitary models over the unitary models in this task, with the gap widening as the system size increases.}}
    \label{fig:gauss}\vspace{-0.5em}
\end{figure}

In this section, we compare the two models $\tilde{\mathcal{E}}_c(\bm{\theta}, \bm{p})$ and $U_c(\bm{\theta})$ on a more practical generative learning task. Specifically, we consider a target distribution created by two overlapping Gaussian distributions, as studied in \cite{liu2018d},
\begin{equation}\label{eq:double_gauss}
    \pi(x) \propto \left( \exp{\left(-\frac{(x-\mu_{1})^2}{2\sigma_g^2}\right)}+ \exp{\left(-\frac{(x-\mu_{2})^2}{2\sigma_g^2}\right)}\right)
\end{equation}
where $\sigma_g$ is the standard deviation of each Gaussian and $\mu_1,\mu_2$ are the two mean values of the distributions. Here, we are assuming discrete (binary) data points $\{x_1,x_2,...,x_M\}$, where $M=2^N$ and $N$ is the number of qubits, to approximate the continuous Gaussian distribution using our variational MBQC. The values of these parameters are chosen similarly to those in  Ref.~\cite{liu2018d}.

We maintain uniformity in our experiments, using the same initialization as in Sec.~\ref{sec:learn_model} as well as the same numbers of qubits ($N=5$ and $N=8$) and corresponding circuit depths ($D=4$ and $D=7$). The chosen sample sizes consist of $8000$ samples for the $N=5$ qubit example and $20,000$ samples for the $N=8$ qubit case. 

In Fig.~\ref{fig:gauss}, a clear learning gap emerges between both models across different system sizes, with the gap widening as the system size increases. Each plot calculates the average loss across eight interactions.

Our empirical results in Secs.~\ref{sec:learn_model} and ~\ref{sec:gauss} show that the generative quantum model based on variational MBQC with probabilistic byproduct correction can provide a learning advantage over the same variational \textit{Ansatz} based on a unitary circuit.\vspace{-0.75em}

\subsection{Comparison between models $U_c$ and $\mathcal{E}_c$}
\label{sec:two_models_compare_alg} 


	In this section, we provide analytic evidence for the observed learning advantage, by comparing $\mathcal{E}_c$ and its unitary counterpart $U_c$ in terms of the set of probability distributions they can generate. In particular, let $\mathcal{Q}^{(N,D)}_{U_c} = \{q_{(\bm{\theta}, \bm{p}=\bm{1})}\}_{\bm{\theta}}$ and  $\mathcal{Q}^{(N,D)}_{\mathcal{E}_c} = \{q_{(\bm{\theta}, \bm{p})}\}_{\bm{\theta}, \bm{p}}$ denote the set of distributions which can be generated using the unitary \textit{Ansatz} $U_c$ and the quantum-channel \textit{Ansatz} $\mathcal{E}_c$ with $N$ qubits and $D$ layers, respectively. Further, let $\Delta^{2^N-1}$ be the $(2^N-1)$-simplex, that is, the set of all $(2^N-1)$ -dimensional distributions. Then, it holds that
	\begin{align}
		\mathcal{Q}^{(N,D)}_{U_c}\subseteq \mathcal{Q}^{(N,D)}_{\mathcal{E}_c}\subseteq \Delta^{2^N-1}
	\end{align}
	where $\mathcal{Q}^{(N,D)}_{U_c}\subseteq \mathcal{Q}^{(N,D)}_{\mathcal{E}_c}$ follows from the fact that ${\mathcal{E}_c}$ reproduces ${U_c}$ when all corrections are applied with probability $p^j_i=1$, and $\mathcal{Q}^{(N,D)}_{\mathcal{E}_c}\subseteq \Delta^{2^N-1}$ holds trivially since $\Delta^{2^N-1}$ contains all probability distributions which can possibly be generated by an $N$-qubit quantum circuit. 
    To better understand the relation between $\mathcal{Q}^{(N,D)}_{U_c}$ and $\mathcal{Q}^{(N,D)}_{\mathcal{E}_c}$ we can view $\mathcal{Q}^{(N,D)}_{U_c}$ as a subset of probability distributions within the simplex $\Delta^{2^N-1}$ and $\mathcal{Q}^{(N,D)}_{\mathcal{E}_c}$ as a convex mixture of distributions in $\mathcal{Q}^{(N,D)}_{U_c}$ (see Fig.~\ref{fig:non_convex}). 

	\begin{figure}[h!]
    \vspace{-0.3cm}
		\centering
		\includegraphics[width=0.9\linewidth]{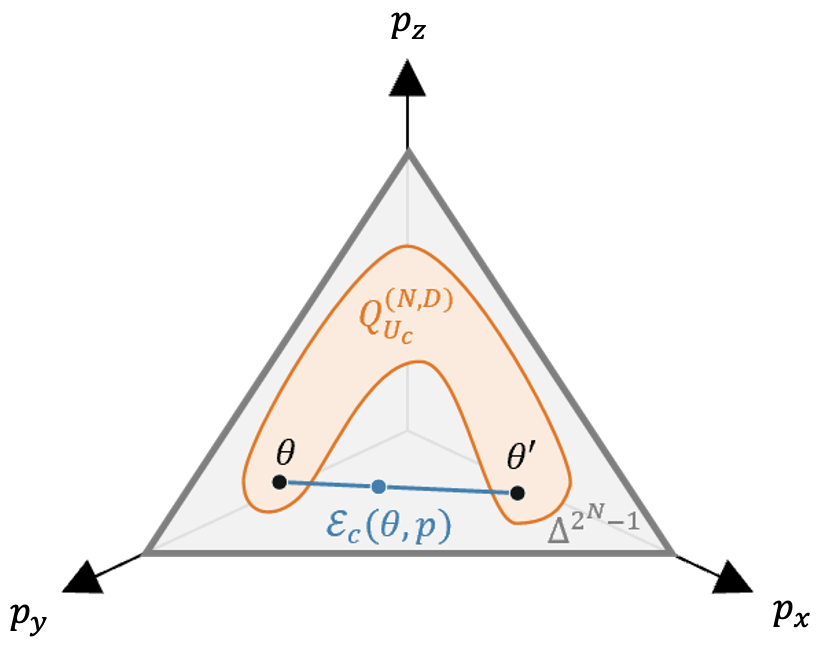}
		\caption{\textbf{Increased expressivity in probabilistic quantum models.} In this qualitative illustration the gray simplex $\Delta^{2^N-1}$ represents the set of all probability distributions for $N$ qubits [here, the $(2^N-1)$-simplex is depicted as a 3-simplex]. The orange area represents a non-convex subset $\mathcal{Q}^{(N,D)}_{U_c}$ which can be generated by the unitary \textit{Ansatz} $U_c$. The blue line represents a convex combination of distributions in $\mathcal{Q}^{(N,D)}_{U_c}$ which can be generated with $\mathcal{E}_c$ but not with $U_c$. The three arrows $\{p_x, p_y , p_z\}$ represent the axes of a three-dimensional Cartesian coordinate space.}\label{fig:non_convex}
	\end{figure}
	
	Then, it also becomes clear that there are two cases where it is impossible to have an expressivity advantage for $\mathcal{Q}^{(N,D)}_{\mathcal{E}_c}$: (i)  if $U_c$ is sufficiently expressive such that $\mathcal{Q}^{(N,D)}_{U_c}= \Delta^{2^N-1}$, and (ii) if $N=1$ because $\mathcal{Q}^{(1,D)}_{U_c}$ forms a convex (one-dimensional) set which makes it impossible to reach new probability distributions by forming convex mixtures. In fact it is necessary that $\mathcal{Q}^{(N,D)}_{U_c}$ is a non-convex set for $\mathcal{Q}^{(N,D)}_{\mathcal{E}_c}$ to have an expressivity advantage (see Fig.~\ref{fig:non_convex} for a qualitative illustration). The following theorem proves that there do indeed exist cases where $\mathcal{Q}^{(N,D)}_{\mathcal{E}_c}$ has an expressivity advantage:
	\begin{theorem}\label{theorem:expressivity}
		There exist a number of qubits $N$ and a number of layers $D$ for which the set of probability distributions  $\mathcal{Q}^{(N,D)}_{\mathcal{E}_c}$ is strictly larger than $\mathcal{Q}^{(N,D)}_{U_c}$ in the sense that in the former set there exists at least one distribution which has an $L^{(1)}$ distance to the closest distribution in the latter set which is bounded away from zero by some finite amount.
	\end{theorem}
	\textit{Proof.} The proof is detailed in Appendix~\ref{app:proof} and proceeds by constructing a distribution which lies in $\mathcal{Q}^{(3,1)}_{\mathcal{E}_c}$ but not in $\mathcal{Q}^{(3,1)}_{U_c}$. In particular, for this distribution, we find that its minimal $L^{(1)}$ distance to $\mathcal{Q}^{(N,D)}_{U_c}$ is lower bounded by $1/10$.

 \subsection{Comparison between models $\tilde{\mathcal{E}}_c$ and $\mathcal{E}_c$}\label{sec:two_models_compare} 
In this section, we demonstrate why the model $\tilde{\mathcal{E}}_c(\bm{\theta},\bm{p})$ in Eq.~\eqref{eq:mbqc_channel_model} should be preferred over the model $\mathcal{E}_c(\bm{\theta},\bm{p})$ in Eq.~\eqref{eq:mbqc_channel_not_model}. Note that the only difference between the two models is the byproduct dependent correction at the end of the computation (which effectively removes the first term in Eq.~\eqref{eq:mbqc_unitary_prop}).

To provide insights into the comparative performances and capabilities of the models $\tilde{\mathcal{E}}_c, \mathcal{E}_c$ and $U_c$, we conduct a series of simulations where all models are trained to approximate a target distribution generated from each other model.

\begin{figure}
    \vspace{2em}
    \centering
    \subfloat[\centering ]{{\includegraphics[width=8cm]{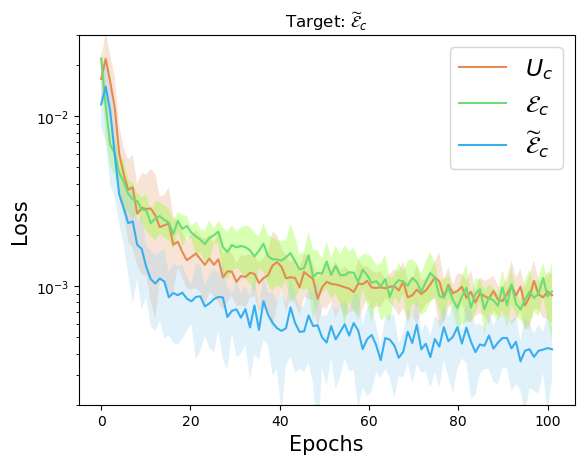} }}%
    \qquad
    \subfloat[\centering ]{{\includegraphics[width=8cm]{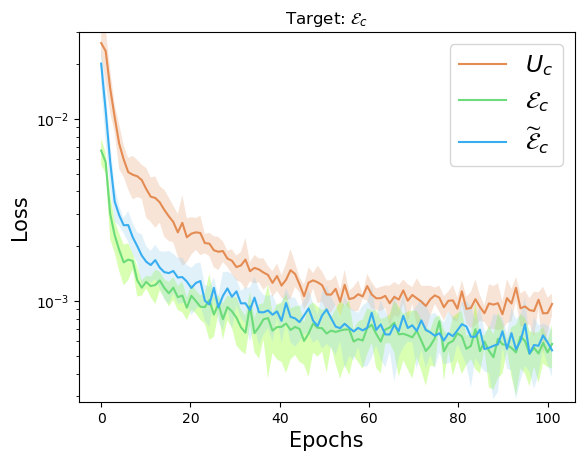} }}%
    \qquad
    \subfloat[\centering ]{{\includegraphics[width=8cm]{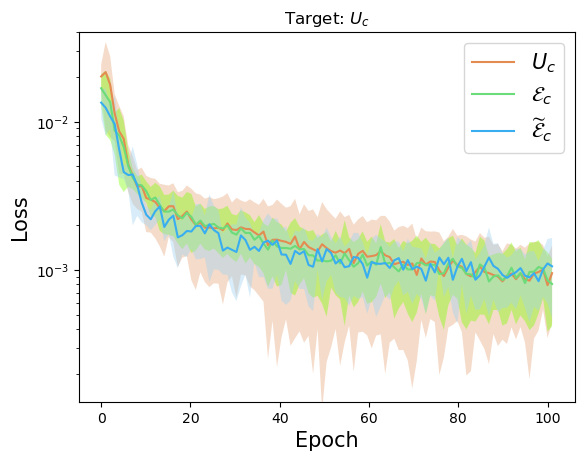} }}%
    \caption{\small{\textbf{Comparison between different learning models.} Here we consider how well each model, $\tilde{\mathcal{E}}_c,\mathcal{E}_c,U_c$, can approximate a target distribution generated from a randomly initialized model based on (a) $\tilde{\mathcal{E}}_c$ (b) $\mathcal{E}_c$ or (c) $U_c$. The $x$-axis represents the number of training steps or epochs the model requires to reach a global optimal point and the $y$-axis represents the variation of the loss function between the model and target distribution, i.e., how close they are. All models were initialized with $N=5$ qubits, depth $D=4$ and trained on $6000$ samples.
    We see that only the learning model based on $\tilde{\mathcal{E}}_c$ can learn all distributions.}}%
    \label{fig:c vs cf}%
\end{figure}

In Fig.~\ref{fig:c vs cf}(a)-(c), all models are trained to approximate different target distributions. Each model is initialized in the same way as in previous sections with $N=5$ and $D=4$. Each training set consists of $6000$ samples and we average over $10$ iterations. In Fig.~\ref{fig:c vs cf}(a) the target distribution is generated from a model $\tilde{\mathcal{E}}_c$, in Fig.~\ref{fig:c vs cf}(b) from a model $\mathcal{E}_c$ and in Fig.~\ref{fig:c vs cf}(c) from a model $U_c$.
We find that only a learning model based on $\tilde{\mathcal{E}}_c$ is able to approximate \emph{all} distributions well, while $\mathcal{E}_c$ fails to learn a distribution based on $\tilde{\mathcal{E}}_c$ better than $U_c$.
This suggests that a learning model based on $\tilde{\mathcal{E}}_c$ is more versatile and expressive than both $U_c$ and $\mathcal{E}_c$ as it can approximate both equally well while also outperforming them in approximating a distribution based on $\tilde{\mathcal{E}}_c$.

Next, we would like to understand why adding Pauli corrections $P_{D+1}(\bm{\tilde{s}})$ (see Eq.~\eqref{eq:mbqc_byproducts}) based on the byproduct occurrence $\bm{\tilde{s}}$ in the circuit, which corresponds to retaining the built-in adaptation of byproducts in MBQC, seems to improve expressivity of the learning model, as suggested by the results above. 

To this end, first note that a byproduct appearing with a certain probability within the circuit of Fig.~\ref{fig:architecture}, can be understood as a noisy single-qubit channel.
Similarly, it is well known that not correcting for byproducts in an MBQC yields a maximally mixed state at the output and hence, a uniform distribution over the output measurements. 
Let us therefore consider byproducts as noise that, eventually, drives the output towards a uniform distribution. 
Our claim is that without the byproduct-dependent correction at the end of the circuit, the corresponding distribution quickly deteriorates towards a uniform distribution with increasing system size.


\begin{figure}[h]
    \centering
    \includegraphics[width=8cm, height=6.5cm]{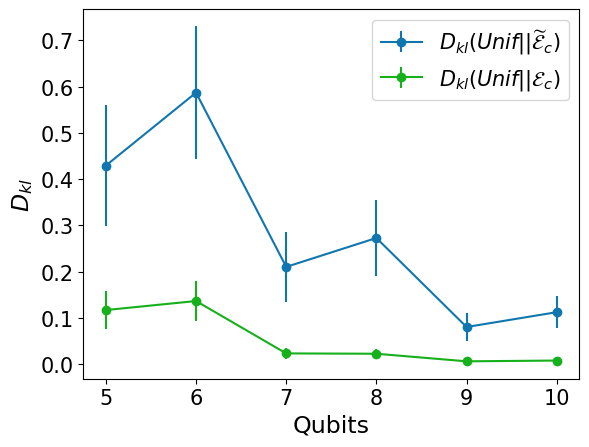}
    \caption{\small{\textbf{$D_{kl}$ of models $\tilde{\mathcal{E}}_c$ and $\mathcal{E}_c$ with the uniform distribution}. 
    Here, we consider how quickly the models $\tilde{\mathcal{E}}_c$ and $\mathcal{E}_c$ deteriorate towards a uniform distribution depending on the system size $N=5,6,...,10$ (and depth $D=N-1$).
    This is done by calculating the Kullback-Leibler Divergence ($D_{kl}$) between the model distribution and a uniform distribution.
    Each model is initialized with random angles $\bm{\theta}$ and, more importantly, random correction probabilities $\bm{p}\in[0.8,1]^{N\times D}$.
     \textit{Blue:} $D_{kl}$ between the uniform distribution and $\tilde{\mathcal{E}}_c$. \textit{Green:}  
    $D_{kl}$ between the uniform distribution and $\mathcal{E}_c$. In both cases, the $D_{kl}$ is averaged over 100 random initializations.
    We find that the blue curve (corresponding to our model with Pauli corrections, $\tilde{\mathcal{E}}_c$) is much slower to deteriorate to a uniform distribution.
    Note that the cyclic behavior most likely stems from the cyclic behavior of the C-QCAs $T_c$ (see Eq.~\eqref{eq:T-func_3}) which has a cycle length that depends on $N$ being odd or even.%
    \label{fig:dkl}}}
\end{figure}

Our claim is confirmed in Fig.~\ref{fig:dkl}, where we plot the Kullback–Leibler Divergence ($D_{kl}$) between the uniform distribution and distributions generated by models $\tilde{\mathcal{E}}_c$ and $\mathcal{E}_c$ across varying system sizes $N=5,6,..,10$ and depths $D=N-1$, averaged over 100 models. Importantly, we initialize the correction probabilities of all models uniformly at random within the range $[0.8, 1]$ (independent of the system size).
The takeaway from this figure is that $D_{kl}(\textrm{Unif}||\mathcal{E}_c)<D_{kl}(\textrm{Unif}||\tilde{\mathcal{E}}_c)$ for all system sizes considered, which indicates that the effect of randomness for generative modeling is significantly enhanced when the propagated byproducts are corrected at the end of the circuit, as it is the case in $\tilde{\mathcal{E}}_c$.

\section{Conclusion and outlook}\label{conclusion}

In this work, we introduce a variational measurement-based quantum computation (MBQC) approach to generative modeling which makes explicit use of the randomness in MBQC that stems from different measurement outcomes. Certain measurement outcomes introduce specific Pauli operators (called \emph{byproducts}) into the computation, which significantly impact the resulting unitary.  Our model features additional learning parameters that govern the probabilities with which we correct for these byproducts such that our generative model effectively becomes a parameterized mixed unitary channel. Importantly, these parameters make use of the classical processing already present in MBQC. In particular, they do not require any additional quantum resources such as additional qubits or quantum gates. Meanwhile, to have the same number of parameters as the channel model, the unitary model would require adding additional qubits to the quantum computation, or increasing the depth of the computation.
However, byproducts have the same effect as noise and quickly deteriorate the result of the computation to a maximally mixed state. To mitigate the detrimental impact of byproducts, we add a byproduct-dependent unitary at the end of the computation. 
This unitary effectively removes the classical, i.e., Clifford, part of byproducts and only retains the non-Clifford effect on the computation. 
To achieve this, we express byproducts as an adaptation of rotational angles and additional Pauli operators at the end of the computation. The latter operators are then corrected in our model while the non-Clifford adaptations to rotations define the effect of randomness in our model.

We demonstrate that our mixed unitary approach to generative modeling outperforms a unitary approach with the same \textit{Ansatz} in two generative learning tasks. Moreover, we can show numerically that retaining the byproduct correction at the end of the computation significantly improves the expressivity and controllability of our model.
In support of our numerical findings, we also provide analytic evidence that learning models based on MBQC with partial byproduct correction can generate probability distributions that a corresponding unitary MBQC cannot.

In this work, we showcase the benefits of using a variational MBQC \textit{Ansatz} for generative modeling. Interestingly, as shown in Refs.~\cite{nautrup2023measurement,stephen2019subsystem}, an MBQC yields a whole family of different ansätze that can be exploited depending on the resource state that is used. 
 Within the regime of intermediate-scale quantum devices, our results suggest that there may be a growing advantage with system size. In fact, we have shown that there exist shallow depth circuits for which the set of distributions that can be generated from a unitary \textit{Ansatz} is smaller than the set of distributions generated by our mixed unitary MBQC \textit{Ansatz} (Sec. \ref{sec:two_models_compare_alg}, Appendix \ref{app:proof}). However, as of now, it is unclear how our methodology will scale to a larger number of qubits and depths, and how the resulting set of distributions will change. We expect that it faces similar challenges with barren plateaus as unitary ansätze~\cite{mcclean2018barren, cerezo2021cost}. To avoid vanishing gradients caused by the use of a sigmoid function, one could explore other methods which map trainable parameters into the interval [0,1].
While we demonstrate our expressiblity advantage in two important testbeds, it remains to be seen whether it holds for other use cases too.
To this end, it would be interesting to explore losses other than maximum mean discrepancy, that do not impose a large classical cost to train the model.
In general, our results motivate further research into enhancing variational quantum algorithms for generative modeling by leveraging MBQC-based ansätze. 

\section{Data availability}
The code used for the numerical simulations is available in
Ref. \cite{Majumder_VMBQC}.
\section{Acknowledgements}
The authors are thankful to V. Dunjko for insightful discussions at the early stages of this project. This research was funded in whole, or in part, by the Austrian Science Fund (FWF) DOI
10.55776/F71. We acknowledge the support of the European
Union (ERC Advanced Grant, QuantAI, No. 101055129). S.J. 
thanks the BMWK (EniQma) for their support. Views and
opinions expressed are however those of the author(s) only
and do not necessarily reflect those of the European Union or
the European Research Council. Neither the European Union
nor the granting authority can be held responsible for them.

\section{Author contributions}
A.M. performed the numerical simulations and analysis. H.J.B., H.P.N., M.K., and S.J. developed the main idea and theory. The implementation has been developed jointly with A.M., M.K. and T.R.. L.J.F., S.J., and H.P.N. worked out the analytical results. H.J.B. conceptualized and supervised the overall project. All authors participated in discussing results and writing and revising the manuscript. 

\section{Competing interest}
The Authors declare no Competing Financial or Non-Financial Interests.
%
%

\bibliographystyle{stdWithTitle}
\bibliography{bib}

\appendix
\section{Cost functions and gradients}\label{cost_detail}
In a manner analogous to the various generative model types (see Sec.~\ref{pqc,gm}), we can also employ both \textit{explicit} and \textit{implicit} loss functions to address different problem scenarios. Explicit loss functions necessitate direct access to both the target and model probability distributions. There are many popular examples of explicit losses, e.g., mean squared error (MSE), KL divergence (KLD) \cite{kullback1951information}, Jenson-Shannon divergence (JSD) \cite{lin1991divergence} and many more. Conversely, implicit loss functions only demand access to samples generated from these two distributions.

An implicit loss is defined as an average over samples drawn from model and target distributions. Implicit losses can be formulated, in general, as 
\begin{equation}
    \mathcal{L}_{\textrm{impl}}(\bm{\theta}):= \mathop{\mathbb{E}}_{x_{1},x_{2},...,x_{M}\sim \{ \pi,q_{\bm{\theta}} \}} g(x_{1},x_{2},...,x_{M})
\end{equation}
Here, $g$ represents a scalar function that exclusively takes the data samples as input and remains independent of the distributions. The expectation is computed across all data samples drawn from either the target, the model, or both distributions.
Here, $g$ can be a scalar function, e.g., a Gaussian kernel, that measures the distance between two data samples drawn from two distributions.

In this paper, we employ a QCBM, an implicit generative model, such that the natural choice for a loss function is an implicit loss. Moreover, authors in Ref.~\cite{rudolph2024trainability} explicitly demonstrated the superiority of implicit losses in training implicit generative models, in contrast to explicit losses that are more appropriate for explicit models. 

Here, Maximum Mean Discrepancy (MMD) \cite{gretton2012k} as an implicit loss function, is defined as follows:

\begin{equation}\label{mmd loss 2}
\begin{aligned}
    \mathcal{L}(\bm{\theta}, \bm{p})= {} & \mathop{\mathbb{E}}_{\substack{x \sim q_{(\bm{\theta}, \bm{p})}\\ y \sim q_{(\bm{\theta}, \bm{p})}}} [K(x,y)]-2\mathop{\mathbb{E}}_{\substack{x \sim q_{(\bm{\theta}, \bm{p})}\\ y \sim \pi}} [K(x,y)] \\
    & +\mathop{\mathbb{E}}_{\substack{x \sim \pi\\ y \sim \pi}} [K(x,y)]
\end{aligned}
\end{equation}
where $q_{(\bm{\theta}, \bm{p})}$ is the distribution from our variational MBQC model, and $\pi$ is the target distribution that the model intends to learn. More specifically,  $\bm{\theta}$ are the variational measurement angles and $\bm{p}$ are the correction probabilities $\{p^j_i\}_{i,j}$ of the Pauli-$Z$ random byproducts, as defined in Eq.~\eqref{eq:prob_variation}, which govern the mixed unitary channel in Eq.~\eqref{eq:mbqc_channel_model}. In this way, we make the degree of non-adaptiveness learnable.

The $K(x,y)$ in Eq.~\eqref{mmd loss 2} is referred to as the kernel function, which is used to introduce a measure of similarity between samples. Here we consider a kernel composed of a mixture of Gaussians, defined as
\begin{equation}\label{kernel}
    K(x,y)=\frac{1}{d}\sum_{k=1}^{d} \exp \left( -\frac{||x-y||^{2}}{2\sigma_{k}}\right)
\end{equation}
where $\sigma_{k} > 0$ is a bandwidth parameter that controls the width of the Gaussian, $d$ is the number of Gaussians mixed, and ||.|| is the $\ell_{2}$-norm between data samples $x$ and $y$. The kernel contributes by providing a continuous measure of distance between the target and model samples \cite{rudolph2024trainability}, and the bandwidth parameters $\sigma_k$ allow us to control the order of magnitude on which samples are considered to be similar. We choose $d=2$ and $\sigma_1, \sigma_2 = 0.5, 4$.
\vspace{8pt}

\vspace{8pt}
\section{Gradient for correction probabilities}\label{gradient correction}
In the following section, we will derive the gradient for the correction probabilities as stated in Eq.~\eqref{gradp}. Our model distribution is denoted as $q_{(\bm{\theta}, \bm{p})}$. As discussed in Sec.~\ref{partially corrected}, we can write the total probability for any measurement outcome $x$ as
\begin{align}\label{total prob}
    q_{(\bm{\theta} , \bm{p})}(x) &= \text{Tr}\left[\mathcal{E}_c(\bm{\theta}, \bm{p})[\rho]\ket{x}\!\bra{x}\right] \\
    &=\sum_{\bm{\tilde{s}}\in\{0,1\}^{N\times D}}
    p(\bm{\tilde{s}}) \langle x | U(\bm{\theta}, \bm{\tilde{s}})\rho U^\dagger(\bm{\theta}, \bm{\tilde{s}})|x\rangle, \nonumber
\end{align}
and similarly for model $\tilde{\mathcal{E}}_c$.

We rewrite the effective byproducts $\tilde{s}_i=(1-c_i)s_i$, $\forall i\in[N \times D]$ in terms of the actual byproducts $s_i\in \{0,1\}$ (that are distributed uniformly and independently), and the random variables $c_i \in \{0,1\}$ that dictate whether we correct for these byproducts or not. From here, we re-parameterize Eq.~\eqref{total prob} in terms of $\bm{c}$ instead of $\bm{\tilde{s}}$:
\begin{align}
    q_{(\bm{\theta} , \bm{p})}(x)  = \sum_{\bm{c}\in\{0,1\}^{N\times D}} p(\bm{c}) q_{(\bm{\theta} , \bm{p})}(x|\bm{c}),
\end{align}
where we absorb the randomness of $\bm{s}$ into
\begin{equation}
q_{(\bm{\theta} , \bm{p})}(x|\bm{c}) =\sum_{\bm{\tilde{s}}\in\{0,1\}^{N\times D}}
    p(\bm{\tilde{s}}|\bm{c}) \langle x | U(\bm{\theta}, \bm{\tilde{s}})\rho U^\dagger(\bm{\theta}, \bm{\tilde{s}})|x\rangle.
\end{equation}
And since we decide to correct each byproduct independently, we can further rewrite
\begin{align}\label{tp2}
    q_{(\bm{\theta} , \bm{p})}(x)  &= \\  & \sum_{\substack{c_{i}\\ \, i\in \{1,\dots , ND\}}} p(c_1)\dots p(c_{ND}) q_{(\bm{\theta} , \bm{p})}(x|c_1,\dots, c_{ND}) \nonumber
\end{align}

Here, $p(c_i) = p_i$ are trainable parameters of our model, and:
\begin{equation}
    c_i =
\begin{cases}
		0  & \mbox{with  probability } 1- p_i,\\
		1 & \mbox{with  probability } p_i.
\end{cases}
\end{equation}

Thus, for any byproduct $k$ under consideration, Eq.~\eqref{tp2} can be written as 

\begin{align}
    &q_{(\bm{\theta} , \bm{p})}(x)= \\ 
    &p_k\cdot \sum_{\substack{c_{i}, \, i\in \{1\dots ND\}, \\ i\neq k}} \left[\prod_{i \ne k }p(c_i) \right]q_{(\bm{\theta} , \bm{p})}(x|c_1,.,1,.,c_{ND}) +  \nonumber
    \\
    & (1-p_k)\sum_{\substack{c_{i}, \, i\in \{1\dots ND\}, \\ i\neq k}} \left[\prod_{i \ne k }p(c_i) \right]q_{(\bm{\theta} , \bm{p})}(x|c_1,.,0,.,c_{ND}) \nonumber
\end{align}

Now, if we take a partial derivative w.r.t. the particular parameter $p_k$, it becomes

\begin{align}
    & \frac{\partial q_{(\bm{\theta} , \bm{p})}(x)}{\partial p_k} = \\ 
    &\sum_{\substack{c_{i}, \, i\in \{1\dots ND\}, \\ i\neq k}}\left[\prod_{i \ne k }p(c_i) \right]q_{(\bm{\theta} , \bm{p})}(x|c_1,.,1,.,c_{ND}) \nonumber 
    \\
    & - \sum_{\substack{c_{i}, \, i\in \{1\dots ND\}, \\ i\neq k}} \left[\prod_{i \ne k }p(c_i) \right]q_{(\bm{\theta} , \bm{p})}(x|c_1,.,0,.,c_{ND}). \nonumber
\end{align}
Applying the law of total probability to all $c_i$ except $c_k$ allows us to marginalize them out, resulting in:
\begin{equation} \label{partial ps}
    \frac{\partial q_{(\bm{\theta} , \bm{p})}(x)}{\partial p_k}=q_{(\bm{\theta} , \bm{p})}(x|c_k=1) - q_{(\bm{\theta} , \bm{p})}(x|c_k=0)
\end{equation}

Considering the MMD loss in Eq.~\eqref{mmd loss}, we can consider only the first two terms as the last one does not depend on the model probabilities $q_{(\bm{\theta} , \bm{p})}$ and the loss can be rewritten as (avoiding the last term) 

\begin{equation}
\begin{aligned}
    L_1(\bm{\theta} , \bm{p})=& \sum_{x ,y } q_{(\bm{\theta}, \bm{p})}(x)q_{(\bm{\theta}, \bm{p})}(y)K(x,y)
    \\
    & -2\sum_{x ,y } q_{(\bm{\theta}, \bm{p})}(x)\pi(y)K(x,y)
\end{aligned}
\end{equation}
Taking a partial derivative of $L_1$, or equivalently $L$ in Eq.~\eqref{mmd loss}, we obtain (using $K(x,y) = K(y,x)$):

\begin{equation}\label{gradl1}
\begin{aligned}
    {}  \frac{\partial L}{\partial p_k} = \ & 2 \sum_{x ,y } \frac{\partial q_{(\bm{\theta}, \bm{p})}(x)}{\partial p_k} q_{(\bm{\theta}, \bm{p})}(y)K(x,y)
    \\
    & -2  \sum_{x ,y } \frac{\partial q_{(\bm{\theta}, \bm{p})}(x)}{\partial p_k}\pi(y)K(x,y) 
\end{aligned}
\end{equation}

Combining this expression with Eq.~\eqref{partial ps}, we get:

\begin{align} \label{Eq:AlmostThere}
    & \frac{\partial L}{\partial p_k} = \\
    &2 \sum_{x ,y } (q_{(\bm{\theta}, \bm{p})}(x | c_k=1)-q_{(\bm{\theta}, \bm{p})}(x|c_k=0))q_{(\bm{\theta}, \bm{p})}(y)K(x,y) \nonumber
    \\
    & -2  \sum_{x ,y } (q_{(\bm{\theta}, \bm{p})}(x|c_k=1)-q_{(\bm{\theta}, \bm{p})}(x | c_k=0)) \pi(y)K(x,y) \nonumber
\end{align}

One can enforce the cases $c_k = 1,0$ by setting $p_k = 1,0$, respectively, and then sample $x$. We introduce the notation $\bm{p}^1$ and $\bm{p}^0$ to refer to the correction probabilities $p_i$, but with $p_k$ replaced by $1$ and $0$, respectively. If we write Eq.~\eqref{Eq:AlmostThere} in the form of expectation values, then it finally becomes

\begin{equation}
\begin{aligned}
    \frac{\partial L}{\partial p_k} &= 2 \mathop{\mathbb{E}}_{\substack{x \sim q_{(\bm{\theta}, \bm{p}^1)}\\ y \sim q_{(\bm{\theta}, \bm{p})}}} [K(x,y)]
     - 2 \mathop{\mathbb{E}}_{\substack{x \sim q_{(\bm{\theta}, \bm{p}^0)}\\ y \sim q_{(\bm{\theta}, \bm{p})}}} [K(x,y)]
    \\
    &\ - 2  \mathop{\mathbb{E}}_{\substack{x \sim q_{(\bm{\theta}, \bm{p}^1)}\\ y \sim \pi}} [K(x,y)]
     + 2  \mathop{\mathbb{E}}_{\substack{x \sim q_{(\bm{\theta}, \bm{p}^0)}\\ y \sim \pi}} [K(x,y)],
\end{aligned}
\end{equation}
which is essentially Eq.~\eqref{gradp}.

In our numerical simulations, we did not use $p_k$ directly as trainable parameters, because of their limited range, but we parametrized them as $p_k = \sigma(\zeta_k)$ with $\sigma(\cdot)$ the sigmoid function and $\zeta_k \in \mathbb R$. This gives an additional inner derivative $\sigma'(\zeta_k)~=~\sigma(\zeta_k)~(1~-~\sigma(\zeta_k))$, which can be dealt with via the chain rule:
\begin{align}
    \frac{\partial L}{\partial \zeta_k} &= \frac{\partial L}{\partial p_k} \frac{\partial p_k}{\partial \zeta_k}\\
    &= \frac{\partial L}{\partial p_k} \sigma(\zeta_k)(1-\sigma(\zeta_k))
\end{align}

To ensure precise gradient estimation, it's essential to accumulate a large number of samples from the quantum circuit. This entails repeating the process multiple times with varying parameter sets. Once the sampling noise is sufficiently mitigated to a predetermined level, this information can then be leveraged to update the parameters on a classical computer. These refined parameters are subsequently fed back into the quantum circuit, facilitating the continuation of the training process.

\section{Proof of Theorem~\ref{theorem:expressivity}}\label{app:proof}
	\textit{Proof.} The first half of the proof proceeds by constructing a distribution which lies in $\mathcal{Q}^{(N,D)}_{\mathcal{E}_c}$ but not in $\mathcal{Q}^{(N,D)}_{U_c}$. Consider $(N,D) = (3,1)$, i.e., $\Delta^{2^N-1}=\Delta^{7}$. Then, along the lines of section II. C in \cite{nakanishi2020sequential}, one can show that the analytical expression of the distribution ${q}^{(3,1)}_{(\bm{\theta, \bm{p}=\bm{1}})}$ resulting from the unitary model $U_c$ must be of the form:
    \begin{equation}
        {q}^{(3,1)}_{(\bm{\theta, \bm{p}=\bm{1}})}(x) = \hat{\bm{c}}(x) \cdot \bigotimes_{i=1}^{3} \begin{pmatrix} \cos(2\theta_i) \\ \sin(2\theta_i) \\ 1 \end{pmatrix}
    \end{equation}
    for every bit-string $x\in\{0,1\}^3$, where $\hat{\bm{c}}(x) \in \mathbb{R}^{27}$ is a vector of coefficients that depends on $x$. More specifically, for the circuit resulting from our MBQC \textit{Ansatz} (i.e., that of Fig.~\ref{fig:architecture} with $(N,D)=(3,1)$), one can derive analytically the coefficients of this expression (in \cite{Majumder_VMBQC}, we provided a Mathematica notebook that performs this derivation) and show that the distribution $\bm{q}^{(3,1)}_{(\bm{\theta, \bm{p}=\bm{1}})}$ can be further simplified to 
	\begin{align}
		\bm{q}^{(3,1)}_{(\bm{\theta, \bm{p}=\bm{1}})}&= \hat{\bm{c}}_0+\hat{\bm{c}}_1\cos2\theta_1\cos2\theta_2\cos2\theta_3+\notag\\
&~~~~\hat{\bm{c}}_2\sin2\theta_1\sin2\theta_2+\hat{\bm{c}}_3\sin2\theta_1\sin2\theta_3+\notag\\
&~~~~\hat{\bm{c}}_4\sin2\theta_2\sin2\theta_3 \label{eq:proof_example}
	\end{align}
	where $\hat{\bm{c}}_0$ points to the center of $\Delta^{7}$ and $\{\hat{\bm{c}}_i\}_{i=1}^4$ span a four-dimensional subset of $\Delta^{7}$ which contains $\mathcal{Q}^{(3,1)}_{U_c}$ and, by convexity, also $\mathcal{Q}^{(3,1)}_{\mathcal{E}_c}$,
\begin{align}
	&\hat{\bm{c}}_0=\frac{1}{8}\left(
		1,
		1,
		1,
		1,
		1,
		1,
		1,
		1
	\right),\\& \hat{\bm{c}}_1=\frac{1}{8}\left(
		-1,
		1,
		1,
		-1,
		1,
		-1,
		-1,
		1
	\right),\\&\hat{\bm{c}}_2=\frac{1}{8}\left(
		1,
		1,
		-1,
		-1,
		-1,
		-1,
		1,
		1
	\right), \\&\hat{\bm{c}}_3=\frac{1}{8}\left(
		1,
		-1,
		1,
		-1,
		-1,
		1,
		-1,
		1
  \right),\\&\hat{\bm{c}}_4=\frac{1}{8}\left(
		1,
		-1,
		-1,
		1,
		1,
		-1,
		-1,
		1
	\right).
\end{align}
The expression in Eq.~\eqref{eq:proof_example} is easy to verify by evaluating the $(3,1)$ model for different parameters $\bm{\theta}$. For convenience we also provide a Mathematica notebook ~\cite{Majumder_VMBQC} which performs this computation.\\
Consider now the following distribution from $\mathcal{Q}^{(3,1)}_{\mathcal{E}_c}$:
\begin{align}
	\bm{q}^{*}&\equiv\bm{q}^{(3,1)}_{(\bm{\theta}=(\pi/8,\pi/8,\pi/8),\bm{p}=(1/2,1/2,1/2))	}\\
	&=\hat{\bm{c}}_0+\hat{\bm{c}}_1\frac{1}{16\sqrt{2}}+\hat{\bm{c}}_2\frac{1}{8}+\hat{\bm{c}}_3\frac{1}{8}+\hat{\bm{c}}_4\frac{1}{8}.
\end{align}
This distribution cannot be represented by $\bm{q}^{(3,1)}_{(\bm{\theta, \bm{p}=\bm{1}})}$ because
\begin{align}
	\bm{q}^{(3,1)}_{(\bm{\theta, \bm{p}=\bm{1}})} = \bm{q}^{*} \label{eq:system_of_equations_proof}
\end{align}
leads to a contraction for all $\bm{\theta}$. To see this, first notice that $\{\hat{\bm{c}}_i\}_{i=0}^4$ forms an orthogonal basis such that Eq.~\eqref{eq:system_of_equations_proof} corresponds to a system of equations with one equation for each $\hat{\bm{c}}_i$. Then, a contradiction is obtained as follows: the equations for $\{\hat{\bm{c}}_i\}_{i=2}^4$ imply $\sin2\theta_1=\sin2\theta_2=\sin2\theta_3=\pm\sqrt{1/8}$, which implies $\cos2\theta_1\cos2\theta_2\cos2\theta_3=\pm \frac{7\sqrt{14}}{32}$, producing a contradiction with the equation for $\hat{\bm{c}}_1$. Thus, $\bm{q}^{*}$ is not contained in $\mathcal{Q}^{(3,1)}_{U_c}$. 

The second half of the proof is a strengthening of the result of the first half of the proof. Specifically we will show that $L^{(1)}$ distance of $\bm{q}^*$ to the closest distribution in $\mathcal{Q}^{(3,1)}_{U_c}$ is bounded away from zero by a finite amount, i.e., we will show that there exists an $\delta>0$ such that 
\begin{align}
    \inf_{\bm{\theta}}D^{(1)}(\bm{q}^*, \bm{q}^{(3,1)}_{(\bm{\theta, \bm{p}=\bm{1}})})> \delta
\end{align}
where $D^{(1)}$ denotes $L^{(1)}$ distance.

Consider the orthogonal basis $\mathcal{C}=\{\bm{c}_0,\bm{c}_1,\bm{c}_2,\bm{c}_3,\bm{c}_4\}$ of the five-dimensional space within which $\mathcal{Q}^{(3,1)}_{\mathcal{E}_c}$ and $\mathcal{Q}^{(3,1)}_{U_c}$ lie. Using vector notation with respect to basis $\mathcal{C}$ we obtain
\begin{align}
	\bm{q}^*=\begin{pmatrix}
		1\\
		1/(16\sqrt{2})\\
		1/8\\
		1/8\\
		1/8
	\end{pmatrix}_{\mathcal{C}},
\end{align}
and
\begin{align}
     \bm{q}^{(3,1)}_{(\bm{\theta, \bm{p}=\bm{1}})}=\begin{pmatrix}
		1\\
		\cos2\theta_1\cos2\theta_2\cos2\theta_3\\
		\sin2\theta_1\sin2\theta_2\\
		\sin2\theta_1\sin2\theta_3\\
		\sin2\theta_2\sin2\theta_3
	\end{pmatrix}_{\mathcal{C}}.
\end{align}
Define
	\begin{align}
		\bm{q}^*(\bm{\epsilon})=\bm{q}^*+\bm\epsilon
	\end{align}
where
\begin{align}
	\bm{\epsilon}=
	\begin{pmatrix}
		0\\
		\epsilon_1\\
		\epsilon_2\\
		\epsilon_3\\
		\epsilon_4
	\end{pmatrix}_{\mathcal{C}}
\end{align}
with $\epsilon_i\in\mathbb{R}$.
Then, we can rewrite the optimization problem as
\begin{align}
    &\inf_{\bm\theta} D^{(1)}(\bm{q}^*,\bm{q}^{(3,1)}_{(\bm{\theta, \bm{p}=\bm{1}})})=\notag\\
    &\quad \inf_{\bm{\epsilon}:D^{(1)}(\bm{q}^*(\bm{\epsilon}),\bm{q}^{(3,1)}_{(\bm{\theta, \bm{p}=\bm{1}})})=0}(|\epsilon_1|+|\epsilon_2|+|\epsilon_3|+|\epsilon_4|).
\end{align}
The condition $D^{(1)}(\bm{q}^*(\bm{\epsilon}),\bm{q}^{(3,1)}_{(\bm{\theta, \bm{p}=\bm{1}})})=0$ leads to a system of four equations corresponding to the four $\theta_i$-dependent coefficients of $\bm{q}^{(3,1)}_{(\bm{\theta, \bm{p}=\bm{1}})}$. The bottom three of these equations can be solved for $\theta_i$ which are then inserted in the first equation, which yields
\begin{align}
	\sqrt{1-\frac{(\frac{1}{8}+\epsilon_2)(\frac{1}{8}+\epsilon_3)}{(\frac{1}{8}+\epsilon_4)}}\sqrt{1-\frac{(\frac{1}{8}+\epsilon_3)(\frac{1}{8}+\epsilon_4)}{(\frac{1}{8}+\epsilon_2)}}\times\notag\\
 ~\sqrt{1-\frac{(\frac{1}{8}+\epsilon_2)(\frac{1}{8}+\epsilon_4)}{(\frac{1}{8}+\epsilon_3)}}=\frac{1}{16\sqrt{2}}+\epsilon_1.
\end{align}
Consider now all $\bm{\epsilon}$ with $|\bm{\epsilon}|=|{\epsilon}_1|+|{\epsilon}_2|+|{\epsilon}_3|+|{\epsilon}_4|\leq r$ such that $0<r<\frac{7}{64}$. Then, a lower bound for the left-hand side (LHS) is obtained by forming a lower bound for each of the three factors independently. Each of the factors is lower bounded by replacing the fraction through an upper bound. To form this upper bound for all $\bm{\epsilon}\leq r<\frac{7}{64}$, first notice that the $\epsilon_i$ in the numerator must be chosen positive, the $\epsilon_i$ in the denominator must be chosen negative, and that all $\epsilon_i$ must be such that $\bm{\epsilon}= r$. Next, notice that the joint weight of the $\epsilon_i$ in the numerator, say $k=\epsilon_i+\epsilon_j$, must be distributed equally in order to maximize the numerator, $\epsilon_i=\epsilon_j=k/2$. The resulting expression can then be shown to be maximal for $k=0$.

After some more algebra, this upper bound is found to be attained by choosing both $\epsilon_i$ in the numerator to be zero and the $\epsilon_i$ in the denominator to be $-r$. Clearly, the RHS is upper bounded by setting $\epsilon_1=r$. This leads to  
\begin{align}
\mathrm{LHS}&\geq\left(1-\frac{1}{64(\frac{1}{8}-r)}\right)^{\frac{3}{2}}\\
\mathrm{RHS}&\leq \frac{1}{16\sqrt{2}}+r.
\end{align}
This is easily shown to be incompatible for sufficiently small $r$. For example, we set $r=\frac{1}{10}$. We then find that \begin{align}
\frac{\mathrm{LHS}}{\mathrm{RHS}}&\geq\frac{15\sqrt{6}}{16+5\sqrt{2}}> 1.
\end{align}
We have thus shown that there exists a finite $\delta$, specifically $\delta=\frac{1}{10}$, such that
\begin{align}
    \inf_{\bm{\theta}}D^{(1)}(\bm{q}^*, \bm{q}^{(3,1)}_{(\bm{\theta, \bm{p}=\bm{1}})})> \delta.
\end{align}
\hfill $\square$

\section{Noisy simulation}
In this section, we provide numerical results for a generative learning task similar to the task in Sec.~\ref{sec:gauss}, but in the presence of incoherent noise. For our experiments, we choose the depolarizing noise model (which includes all Pauli-$X, Y, Z$ errors with equal probabilities) that acts on each qubit at the end of each layer in the circuit (Fig. ~\ref{fig:architecture}).

The action of the depolarizing channel can be described as 

\begin{equation}
     \epsilon [\rho] = (1-p_n)\rho + p_n \frac{\mathbb{I}}{2^N}
\end{equation}

where $\rho$ is the initial state of the system and $\mathbb{I}/2^N$ is the maximally mixed state with $N$ being the total number of qubits. Here, $p_n$ refers to the noise strength.

To demonstrate the performance of our VMBQC model in the presence of depolarizing noise, we choose the $\mathcal{E}_c(\bm{\theta},\bm{p})$ model (where we do not correct the intermediate wrong measurement outcomes) with 5-qubits and 4-layers that acts as the mixed unitary channel. The other model,  $\tilde{\mathcal{E}}_c(\bm{\theta},\bm{p})$ (where we correct the wrong outcomes), is expected to behave similarly in the presence of noise.

\begin{figure}[H]
    \centering
    \includegraphics[width=8cm, height=6.5cm]{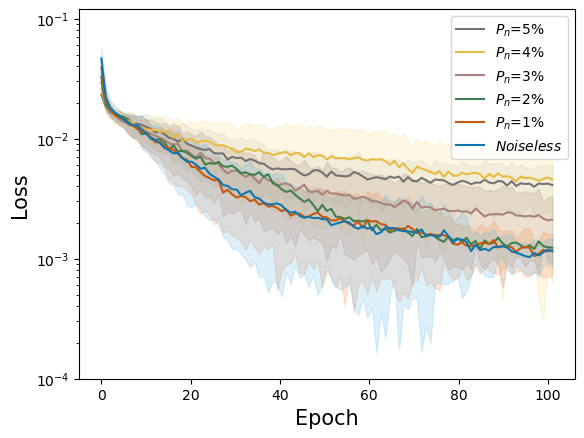}
    \caption{\small{\textbf{ Variational MBQC learning performance on a double Gaussian under \textit{depolarizing} noise.} Here we showcase the behavior of the $\mathcal{E}_c(\bm{\theta},\bm{p})$ model under the depolarizing noise with varying strength. The model is trained, each time, with 6000 training samples with $N$=5 qubits and $D$=4 layers. The $x$-axis represents the time steps (epochs), the model takes to reach an optimal value while the $y$-axis shows the loss values between the model and target distributions.
    }}
    \label{fig:noisy-sim-Dep}\vspace{-0.5em}
\end{figure}

As expected, we observe, in Fig.~\ref{fig:noisy-sim-Dep}, a decrease in the model's performance with increasing noise. For small noise levels, $<2\%$, the performance of the model barely suffers. However, the performance will suffer more for a larger number of qubits and depths.

We also perform noisy simulations to compare the performances of the unitary and channel models in Fig.\ref{fig:noisy-sim-unit_vs_chann}, under varying noise strength. Here we observe that the channel model performs moderately better than the unitary model at a low noise regime.

\begin{figure}[H]
    \centering
    \includegraphics[width=8cm, height=6.5cm]{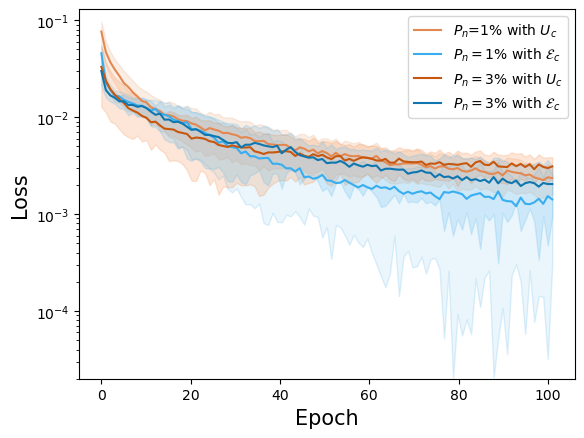}
    \caption{\small{\textbf{Comparison of unitary and channel models under \textit{depolarizing} noise.} Here we showcase the behavior of the $\mathcal{E}_c(\bm{\theta},\bm{p})$ and $U_c$ models under the depolarizing noise with varying strength. The model is trained, each time, with 6000 training samples with $N$=5 qubits and $D$=4 layers. The $x$-axis represents the time steps (epochs), the model takes to reach an optimal value while the $y$-axis shows the loss values between the model and target distributions.
    }}
    \label{fig:noisy-sim-unit_vs_chann}\vspace{-0.5em}
\end{figure}
\end{document}

%% file: definitions.tex
\usepackage{algorithm,algpseudocode}

\usepackage{enumitem}

\usepackage{hyperref} 
\usepackage{jabbrv}

\usepackage{enumitem}
\let\olddesc\description
\def\description{\olddesc\setlist[itemize]{leftmargin=*,labelindent=-12pt}}

\usepackage{lineno}

\usepackage{xr}

\usepackage[english]{babel}

\usepackage{graphicx}
\usepackage{caption}
\usepackage{subcaption}

\usepackage{xcolor}

\usepackage{tcolorbox}
\tcbuselibrary{breakable}
\tcbuselibrary{skins}
\tcbset{enhanced, colback=white, arc=0mm, outer arc=0mm, fonttitle=\bfseries, fontupper=\small}
\newtcolorbox[auto counter]{examplebox}[2][]{%
title=Box~\thetcbcounter: #2,#1}

\usepackage{amsmath}
\usepackage{amsthm}
\usepackage{amssymb}
\usepackage{mathtools}
\usepackage{thmtools}

\usepackage{bm}  
\usepackage{bbm}
\usepackage{dsfont}
\usepackage{cases}
\usepackage{nicefrac} 
\usepackage{xfrac}

\usepackage{multicol}

\newtheorem{theorem}{Theorem}

\newcommand{\ket}[1]{\left| #1 \right>} 
\newcommand{\bra}[1]{\left< #1 \right|} 

\let\epsilon\relax
\def\epsilon{\varepsilon}
\let\phi\relax
\def\phi{\varphi}

\usepackage[usenames,dvipsnames]{xcolor}
 \definecolor{arun}{rgb}{.2,0.5,.5}
   
 \definecolor{tina}{rgb}{0.7,0,0.9}
 	
\definecolor{fuchsia}{rgb}{1.0, 0.0, 1.0}